\documentstyle[aps]{revtex}

\begin{document}
\draft
\title{Superradiant light scattering and grating formation in cold atomic vapours}
\author{R.\ Bonifacio$^{a}$, B.W.J.\ McNeil$^{b}$, 
N.\ Piovella$^{a}$, G.R.M. Robb$^{b}$\\
$^{a}$ {\it Dipartimento di Fisica, Universit\`a Degli Studi di 
Milano,\\
INFN and INFM, Sezione di Milano, Via Celoria 16, Milano I-20133,Italy\\
$^{b}$ Department of Physics and Applied Physics, University of 
Strathclyde,\\ 
Glasgow G4 0NG, Scotland, UK}}
\maketitle

\begin{abstract}

A semi-classical theory of coherent light scattering from an elongated sample of 
cold atoms exposed to an off-resonant laser beam is presented. The model, which is a direct 
extension of that of the collective atomic recoil laser (CARL), describes the emission of 
two superradiant pulses along the sample's major axis simultaneous with the formation of a 
bidimensional atomic grating inside the sample. It provides a simple physical picture of the 
recent observation of collective light scattering from a Bose-Einstein condensate 
[S. Inouye {\it et al.}, Science {\bf 285}, 571 (1999)]. 
In addition, the model provides an analytical description of the temporal evolution of the 
scattered light intensity which shows good quantitative agreement with the experimental 
results of Inouye {\em et. al}.
\end{abstract}

\section{Introduction}
Recent experiments by S. Inouye {\it et al.} at MIT \cite{MIT} have demonstrated the
formation of atomic matter waves in a cigar-shaped Bose-Einstein Condensate (BEC) 
pumped by an off-resonant laser beam, together with highly directional scattering of light
along the major axis of the condensate. This emission has been interpreted as superradiant
Rayleigh scattering, and some theoretical work describing this experiment has been
recently published \cite{Meystre1,Ozgur}. In particular, the work of Moore and 
Meystre \cite{Meystre1} describes the 
Rayleigh scattering in a BEC using a model which extends the Collective Atomic Recoil 
Laser (CARL) model originally proposed by Bonifacio et 
al. \cite{CARL1,CARL2,CARL3,CARL4} to include a quantum-mechanical description of the 
centre-of-mass motion of the atoms in the condensate \cite{Meystre2,Meystre3}. 
The conclusions of ref.\cite{Meystre2} were that the original CARL theory, which treats the 
atomic centre-of-mass motion classically, fails when the temperature of the atomic sample 
is below the recoil temperature $T_R=\hbar\omega_r/k_B$, where $\omega_r=\hbar |\vec q|^2/2m$ 
is the recoil frequency, $m$ is the atomic mass, $\vec q=\vec k-\vec k_s$ is the difference 
between the pump and scattered wavevectors and $k_B$ is Boltzmann's constant. 
However, the cubic dispersion relation derived in ref.~\cite{Meystre2} reduces to that of
the original semiclassical CARL model for large atomic densities:
More specifically, the quantum corrections to the classical motion are negligibly small 
when the CARL parameter $\rho$ in the free electron laser (FEL) limit \cite{CARL3,CARL4}, 
roughly interpreted as the average number of photons scattered per atom, is greater than one. 
This suggests that a fully quantum-mechanical description of the atomic centre-of-mass motion 
may not be necessary in order to describe the main experimental results of ref.\cite{MIT}, 
i.e.  the temporal evolution of the scattered light intensity and the spatial grating in the 
condensate. We are aware that a semiclassical theory is necessarily limited in its description 
of  the radiation statistics and the quantum degenerate nature of the condensate, which 
require a full quantum analysis. Nevertheless, we consider the semiclassical approach useful 
in order to give an intuitive description of the physical mechanism underlying the observed 
effects. We stress however that in spite of its simplicity, the semiclassical model produces 
good quantitative agreement with the experimental results of ref.\cite{MIT}.

\section{Model}
The model described is bidimensional and semiclassical. We represent the cigar-shaped 
atomic sample as an ellipsoid with length $L$ and diameter $W$, where $L\gg W$ as shown in Fig.\ref{fig1}. 
The sample is exposed to a classical plane wave radiation electric field 
$\vec E_0(y,t)=\hat x {\cal E}_0 e^{ik(y-ct)}+c.c.$, 
polarised along the $\hat x$ axis and incident along the axis $\hat y$, with ${\cal E}_0$ 
real and constant and where 
$k=\omega/c$. We assume that the scattered radiation consists of two radiation pulses 
propagating along the $\hat z$ axis, with electric fields polarised as the incident field:
\begin{equation}
\vec E(z,t)=\hat x[
{\cal E}_1(z,t) e^{ ik(z-ct)}+
{\cal E}_2(z,t) e^{-ik(z+ct)}+c.c],
\label{field}
\end{equation}
where ${\cal E}_{1,2}(z,t)$ are slowly varying complex amplitudes. The dominance of the scattering   
along  the $\pm \hat z$ axis over that in other directions is due to the geometry of the atomic sample.

The atomic sample is described as a collisionless gas of atoms, each with two internal 
energy levels. The internal evolution of each atom is described by the density matrix 
elements $\rho_{mn}$ ($m,n=1,2$) for the lower, (1), and upper, (2), levels. 
The off-diagonal elements $\rho_{12}=\rho_{21}^*$ describe the dipole moment induced by 
the radiation fields via the relation $\vec d=\hat x \mu(\rho_{12}+c.c)$, where $\mu$ is 
the dipole matrix element. The diagonal elements $\rho_{11}$ and $\rho_{22}$ describe the 
probability of an atom being in the lower or in the upper level, respectively. 
The off-diagonal elements may be described conveniently as a sum of three polarisation waves:
\begin{equation}
\rho_{12}=S_0 e^{ ik(y-ct)}+ S_1 e^{ ik(z-ct)}+S_2 e^{-ik(z+ct)}.
\label{polarisation}
\end{equation}
The dipole moment of each atom contributes to the macroscopic polarisation of the atomic 
sample described by $\vec P=n(\vec x)\vec d$, where $n(\vec x)$ is the atomic density. 
This polarisation is a source for the radiation field via Maxwell's wave equation which 
yields, in the usual Slowly Varying Envelope Approximation (SVEA),
\begin{equation}
\left(\frac{\partial {\cal E}_1}{\partial t}
    +c\frac{\partial {\cal E}_1}{\partial z}\right)e^{ikz}+
\left(\frac{\partial {\cal E}_2}{\partial t}
    -c\frac{\partial {\cal E}_2}{\partial z}\right)e^{-ikz}=
\frac{i\omega\mu}{2\epsilon_0}n(\vec x)
\left\{S_0e^{iky}+S_1e^{ikz}+S_2e^{-ikz}\right\},
\label{maxwell}
\end{equation} 
where we have neglected the terms proportional to $e^{\pm2i\omega t}$.
We assume that the atomic sample can be described as a collection of $N$ point particles
with positions $\vec x_j$, so that $n(\vec x)=\sum_{j=1}^N \delta^{(3)}(\vec x - \vec x_j)$. 
Multiplying both sides by $e^{\mp ikz}$ and integrating over the $\hat z$ axis from 
$z-\Delta z/2$ to $z+\Delta z/2$, where $\Delta z=\lambda/2$, Eq.(\ref{maxwell}) yields
\begin{equation}
\left(\frac{\partial {\cal E}_{1,2}}{\partial t}
 \pm c\frac{\partial {\cal E}_{1,2}}{\partial z}\right)\Delta z=
\frac{i\omega\mu}{2\epsilon_0}
\left\{S_0e^{ik(y\mp z_j)}+S_{1,2}+S_{2,1}e^{\mp 2ikz_j}\right\}
\delta(x-x_j)\delta(y-y_j),
\label{max2}
\end{equation} 
where the upper sign corresponds to the first subscript and we have assumed the field 
amplitudes ${\cal E}_{1,2}$ are spatially slowly varying over $\Delta z$. 
Assuming also that ${\cal E}_{1,2}$ are independent of $x$ and $y$, we can integrate on the plane 
$(x,y)$ over the section $A=\pi W^2/4$ of the condensate, so that Eq.(\ref{max2}) becomes
\begin{equation}
\left(\frac{\partial {\cal E}_{1,2}}{\partial t}
 \pm c\frac{\partial {\cal E}_{1,2}}{\partial z}\right)=
\frac{i\omega\mu\bar n}{2\epsilon_0}
\left\langle S_0e^{ik(y\mp z)}+S_{1,2}+S_{2,1}e^{\mp 2ikz}\right\rangle,
\label{max3}
\end{equation} 
where $\bar n=N/A\Delta z$ is the average density and $\langle..\rangle=(1/N)\sum_{j=1}^N(..)_j$.

In this model the atomic centre-of mass motion is treated classically, with each atom 
described as a point particle with a given position and momentum. 
The radiation fields drive the centre-of-mass motion of the atoms via the force 
\[
\vec F=\left( 0 \;\; , \;\; \vec d \cdot \frac{\partial (\vec E_0 + \vec E)}{\partial y} \;\; , \;\;\vec d \cdot \frac{\partial (\vec E_0 + \vec E)}{\partial z} \right) . 
\]
Neglecting the fast-varying temporal terms,  the equations for the atomic velocity components are:
\begin{eqnarray}
m\frac{dv_y}{dt}&=&ik\mu {\cal E}_0\left[
S_0^*+S_1^*e^{ik(y-z)}+S_2^*e^{ik(y+z)}-c.c
\right]\label{vy}\\
m\frac{dv_z}{dt}&=&ik\mu\left\{
 S_1^*{\cal E}_1
-S_2^*{\cal E}_2
+S_2^*{\cal E}_1e^{ 2ikz}
-S_1^*{\cal E}_2e^{-2ikz}
+S_0^*\left[
             {\cal E}_1e^{ik(z-y)}
            -{\cal E}_2e^{-ik(y+z)}
      \right]-c.c.
\right\}.
\label{vz}
\end{eqnarray}
We assume that the detuning $\delta=\omega-\omega_a$ between the optical fields
and the atomic resonance is much larger than the natural linewidth of the atomic 
transition, $\gamma$, so that the atoms always remain in their lower internal energy states
($\rho_{11}\approx 1$ and $\rho_{22}\approx 0$). Moreover, assuming that the scattering 
time scale is much longer than the relaxation time $\gamma^{-1}$, we can adiabatically 
eliminate the atomic polarisations, i.e. 
$S_k=i(\mu/\hbar){\cal E}_k/(\gamma+i\delta)\approx \Omega_k/2\delta$,
where $\Omega_k=2\mu {\cal E}_k/\hbar$, $|\Omega_k|$ is the Rabi frequency for the field 
$k$ and $k=0,1,2$. With these approximations, Eqs.(\ref{vy}) and (\ref{vz}) yield:
\begin{eqnarray}
m\frac{dv_y}{dt}&=&-i\hbar k(\Omega_0/4\delta)
\left[\Omega_1 e^{ik(z-y)}-\Omega_2^* e^{ik(z+y)}-c.c\right]\label{vy:2}\\
m\frac{dv_z}{dt}&=&i\hbar k(\Omega_0/4\delta)
\left[\Omega_1 e^{ik(z-y)}+\Omega_2^* e^{ik(z+y)}-c.c\right]
+i(\hbar k/2\delta)
\left[\Omega_1\Omega_2^* e^{2ikz}-c.c\right].
\label{vz:2}
\end{eqnarray}
It is seen that the interference between the pump and the scattered fields forms two
bidimensional periodic potentials 
$V_{1,2}(y,z)\propto|{\cal E}_0{\cal E}_{1,2}|\cos[k(z\mp y)\pm\phi_{1,2}]$ in the plane 
$(\hat y,\hat z)$, where $\phi_{1,2}$ are the phases of the complex amplitudes 
${\cal E}_{1,2}$. A weaker 1D potential 
$V_3(z)\propto|{\cal E}_1{\cal E}_{2}|\cos[2kz+\phi_1-\phi_2]$ forms along the $\hat z$ axis 
due to the interference of the two counterpropagating scattered fields. 
If the pump intensity is large enough, we can assume ${\cal E}_0\gg {\cal E}_{1,2}$ and 
neglect the ponderomotive potential $V_3$. Then, Eqs.(\ref{vy:2}),(\ref{vz:2}) and (\ref{max3})
can be conveniently written in the following dimensionless form \cite{CARL2}:
\begin{eqnarray}
\frac{d\theta_{1,2}}{d\overline t}&=&p_{1,2},\label{F1}\\
\frac{dp_{1,2}}{d\overline t}&=&\mp\left[A_{1,2}e^{\pm i\theta_{1,2}}+c.c\right],
\label{F2}\\
\frac{\partial A_{1,2}}{\partial \overline t}&\pm&
\frac{\partial A_{1,2}}{\partial\overline z}=\langle e^{\mp i\theta_{1,2}}\rangle
\label{F3}
\end{eqnarray}
where $\theta_{1,2}=k(z\mp y)$, $p_{1,2}=(m/\hbar k\rho)(v_z\mp v_y)$ and 
$A_{1,2}=-2i(\epsilon_0/\hbar\omega\overline n \rho)^{1/2}{\cal E}_{1,2}$ are 
scaled atomic position, atomic momentum and field amplitude variables respectively.
The dimensionless time and space coordinates, $\overline t=\omega_r\rho t$ and 
$\overline z=\omega_r\rho z/c$, are scaled in terms of the collective recoil bandwidth, 
$\rho\omega_r$, where $\omega_r=\hbar k^2/m$ is the single-atom recoil frequency and 
$\rho=(\Omega_0/2\delta)^{2/3}
(\omega\mu^2\overline n/\epsilon_0\omega_r^2\hbar)^{1/3}$ is the dimensionless
CARL parameter \cite{CARL3,CARL4}. At $\overline t=0$, the atoms are assumed to be randomly 
distributed in position and have zero momentum, and the amplitudes of the scattered fields 
are set to zero.

\section{Analysis}
In this simple model the two scattered fields are uncoupled and symmetric.
For each field (1,2) individually, Eqs.(\ref{F1})-(\ref{F3}) are formally identical to 
those which describe pulse propagation in a high gain free electron laser (FEL) \cite{CARL4}. 
It is already known that they admit a self-similar solution of the form 
$A_{1,2}(\overline z,\overline t)=\pm\overline z{\cal A}(u)$, where 
$u=\sqrt{|\overline z|}(\overline t\mp\overline z)$ and ${\cal A}(u)$ is the solution of a 
set of ordinary differential equations \cite{FEL}. This self-similar solution describes the 
superradiant emission of radiation pulses whose duration decreases in proportion to the fourth 
root of the peak intensity. The pulse shape can be approximated by a hyperbolic secant function,
followed by some non-linear `ringing', similar to that which occurs in superfluorescence from 
inverted two-level atoms \cite{SF}. 

A simpler model can be obtained by approximating the spatial derivative in the field equation 
(\ref{F3}) by a damping term\cite{BC} i.e.
\begin{equation}
\frac{dA_{1,2}}{d\overline t}=\langle e^{\mp i\theta_{1,2}}\rangle
-\kappa A_{1,2}
\label{F4}
\end{equation}
where $\kappa=c/2\omega_r\rho L$ and $L/c$ is the transit time of the photon along the major 
axis of the condensate. In this approximation, the finite interaction time due to the escape 
of radiation from the atomic sample is represented by an incoherent decay of the field 
amplitude in the sample at a rate $c/2L$, half the inverse of the radiation `lifetime' 
in the atomic sample.
A more general treatment where the radiation is scattered in a direction making an 
angle $\psi$ with respect to the $\hat z$ axis should give 
$\kappa\approx (c/2\omega_r\rho)[|\sin{\psi}|/W+|\cos{\psi}|/L]$ \cite{SF2}. 
As $L\gg W$, the radiation is least strongly damped along the major axis of the 
sample. 

An approximate solution to eqs.(\ref{F1}),(\ref{F2}) and (\ref{F4}) can be found assuming 
$\kappa\gg 1$ and adiabatically eliminating the field variables, i.e. 
$A_{1,2}\approx \kappa^{-1}\langle\exp[\mp i\theta_{1,2}]\rangle$. In this limit, the 
rate of change of the average scaled momentum is 
$(d/d\overline t)\langle p_{1,2}\rangle=\mp 2\kappa|A_{1,2}|^2$.  
A third-order analysis of the equations in the mean-field limit (i.e. with radiation 
propagation modelled by the damping term) gives the following approximate solution:
\begin{equation}
|A_{1,2}|^2\approx\frac{1}{2\kappa^2}{\rm sech}^2
\left[(\overline t-\overline t_D)/\sqrt{2\kappa}\right]
\label{sech}
\end{equation}
and
\begin{equation}
\langle p_{1,2}\rangle \approx\mp\sqrt{\frac{2}{\kappa}}
\left\{
1+{\rm tanh}\left[(\overline t-\overline t_D)/\sqrt{2\kappa}\right]
\right\},
\label{tanh}
\end{equation}
where $\overline t_D=-\sqrt{2 \kappa} \ln{(|b_0|/\sqrt 2)}$ is the delay time of the peak 
and $b_0=\langle\exp[\mp i\theta_{1,2}({\overline t}=0)]\rangle$ is the initial bunching, 
which can be assumed to be $\sim 1/\sqrt{N}$ for a condensate of $N$ atoms. 
In the linear regime the exponential gain is 
\begin{equation}
G=\omega_r\rho\sqrt{2/\kappa}=(3\gamma/\delta) \sqrt{(2I_0 N/m\omega)(\lambda^2/A)},
\label{gain}
\end{equation} 
whereas the peak value of the scattered intensity is 
\begin{equation}
I_{peak}=(\gamma/\delta)^2[(3/2\pi)(\lambda^2/A)N]^2 I_0,
\label{peak}
\end{equation}
where $I_0=2c\epsilon_0|{\cal E}_0|^2$ is the pump intensity, $A$ is the cross sectional area of the 
condensate and $\gamma=\mu^2k^3/6\pi\hbar\epsilon_0$ is the natural decay rate of the atomic
transition. 

\section{Comparison with the MIT experiment}
In the MIT experiment, a sodium BEC was exposed to a single off-resonant laser pulse
red-detuned by $\delta/2\pi=1.7$ GHz from the $3S_{1/2}\rightarrow 3P_{3/2}$ transition, with
$\lambda=0.589 \mbox{$\mu$m}$ and natural width $\gamma=0.31 \times 10^8 $ s$^{-1}$. 
The recoil frequency is $\omega_r=3 \times 10^5$ s$^{-1}$.
We assume that the condensate had a diameter of $20\ \mbox{$\mu$ m}$ and a 
length of $200 \mbox{$\mu$m}$, approximately $N=5\times 10^5$ atoms participate in the emission 
of a scattered radiation pulse. The dimensionless parameters are 
$\rho\approx 44 \times I_0^{1/3}$ and $\kappa\approx 5.5 \times 10^4 \times I_0^{-1/3}$, 
where $I_0$ is the pump intensity in mW/cm$^2$. As $I_0 >1$ and consequently $\rho\gg 1$, 
the results of ref.~\cite{Meystre2} indicate that quantum effects due to atomic diffraction 
should be negligibly small for this experiment, even though $T \ll T_R$.
As $\kappa\gg 1$, the atoms emit two superradiant pulses along the major axis of the 
condensate. The gain is approximately $G\approx 82 \times \sqrt{I_0}$, where $G$ is given in 
ms$^{-1}$, whereas the peak occurs after a time $t_D=\ln(2 N)/G \approx (170/\sqrt{I_0}) \mbox{$\mu$s}$, 
in good agreement with the measured values of ref.\cite{MIT}. Furthermore, 
from Eq.(\ref{tanh}) the modulus of the average atomic velocity is 
$v=(\hbar k/m)\rho|\langle p_{1,2}\rangle|
\approx(\lambda/2\pi)G \approx 0.7\sqrt{I_0} \mbox{cm s}^{-1}$. 
Fig.~\ref{fig2} shows the temporal evolution of the main peak of the scattered intensity, 
as given by the approximate formula (\ref{sech}), 
for the parameters of the MIT experiment and three different values of the incident intensity,
3.8 mW/cm$^2$ (solid line), 2.4 mW/cm$^2$ (dashed line) and 1.4 mW/cm$^2$ (dotted line).

In addition to the temporal evolution of the scattered radiation pulses, there are other 
predictions of  this semiclassical model which are consistent with the results of the MIT 
experiment: Firstly, superradiance is observable only if the Doppler broadening of the atomic 
resonance is sufficiently small that $\sigma t_{SR}\ll 1$, where $\sigma$ is the rms spread of 
the gaussian spectral distribution and $t_{SR}\sim 1/G$ is the superradiant time \cite{SF}. 
The observed spectral width of the Bragg resonance of the BEC of approximately 
$5$ kHz \cite{MIT2} (corresponding to a velocity spread of few mm s$^-1$) yields 
$\sigma t_{SR}\sim 0.16 \times I_0^{-1/2}$. We observe that, using  
$\sigma=k(k_BT/m)^{1/2}$, a temperature of only $1\mu K$ (approximately the BEC transition 
temperature for the MIT experiment) would increase the frequency spread by a factor of $15$, 
enough to destroy the superradiant emission. This explains why superradiant emission was 
observed only at the extremely low temperatures below the threshold for Bose-Einstein 
condensation \cite{MIT}.
Secondly, superradiant emission parallel and antiparallel to the $\hat z$ axis induces an 
average atomic velocity $\vec v_{1,2}\approx (G/k)[\hat y\mp\hat z]$, respectively at 45$^0$ 
degree with respect to the negative (positive) direction of the $\hat z$ axis, 
as observed in the MIT experiment. We assume the existence of two distinct families of atoms 
interacting with the two independent superradiant pulses $A_1$ and $A_2$. However different 
orders of atomic velocity, i.e. $\vec v_{m,n}=m\vec v_1+n\vec v_2$, 
with $m,n$ integers, have also been observed. More precisely, the orders $(2,0)$, $(-1,1)$, 
$(0,2)$, $(2,1)$ and $(1,2)$ other than the usual $(1,0)$ and $(0,1)$, have been clearly 
observed in the experiment after increasing the exposure time to the laser source and 
letting the atomic cloud expand ballistically after the interaction \cite{MIT}. 
The formation of this momentum grating can be explained as a sequential superradiant 
scattering process in which the atoms emit $m$ pulses along the positive $\hat z$ and $n$ 
pulses along the negative $\hat z$ axis, acquiring a total recoil velocity $\vec v_{m,n}$. 
The extremely narrow resonance line allows the atoms to emit up to three sequential 
superradiantly scattered pulses before $\sigma t \sim 1$, which is consistent with the 
observation of the atomic momentum distribution.

\section{Conclusions}
In conclusion, we have presented a semiclassical model describing the superradiant Rayleigh 
scattering from a Bose-Einstein condensate observed in Ref.~\cite{MIT}. The model is much 
simpler than those previously used to explain the results of \cite{MIT} as the atomic 
centre-of-mass motion is treated classically. 
The evolution of the scattered intensity and the atomic motion due to recoil as calculated 
from this simple model are in quantitative agreement with the experimental results. The fact 
that quantum centre-of-mass effects such as atomic diffraction are negligible is a 
consequence of the high density of the condensate. In our model the BEC is essentially 
described as a collisionless Doppler-free atomic gas. The results presented here suggest 
that together with its high density, the most important property of the condensate with 
regard to superradiant light scattering is its very low temperature rather than its quantum 
degenerate nature. In this respect the situation is similar to that of ultraslow propagation 
of light in a BEC \cite{Hau}. Subsequent observations of ultraslow propagation in a hot vapour 
\cite{Scully} demonstrated that the significant property of the BEC was that it was a 
Doppler-free optical medium rather than a quantum degenerate one.

\acknowledgements
The authors would like to thank the Royal Society of Edinburgh for support of G.R.M.R. and 
the EPSRC  for support of  B.M$^{\rm c}$N.

\begin{figure}
\caption{The geometry of the scattering esperiment. The filled elipsoid, representing the
atomic condensate with dimensions $L$ and $W$, is illuminated with a single off-resonant laser 
beam of electric field $E_0$ polarised along the $\hat x$ axis and propagating along the 
$\pm \hat z$ axis. The geometry favours the emission of the oppositely directed superradiant 
pulses $E_1$ and $E_2$ along the major axis of the condensate.}
\label{fig1}
\end{figure}

\begin{figure}
\caption{Temporal evolution of the main peak of the scattered intensity as given by the
approximate formula (14), for the parameters of the MIT experiment and three different values 
of the incident intensity, 3.8 (solid line), 2.4 (dashed line) and 1.4 (dotted line) mW/cm$^2$.}
\label{fig2}
\end{figure}

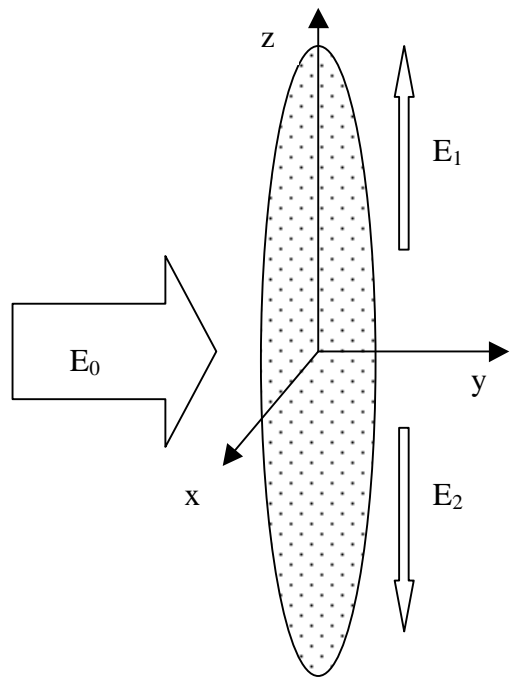
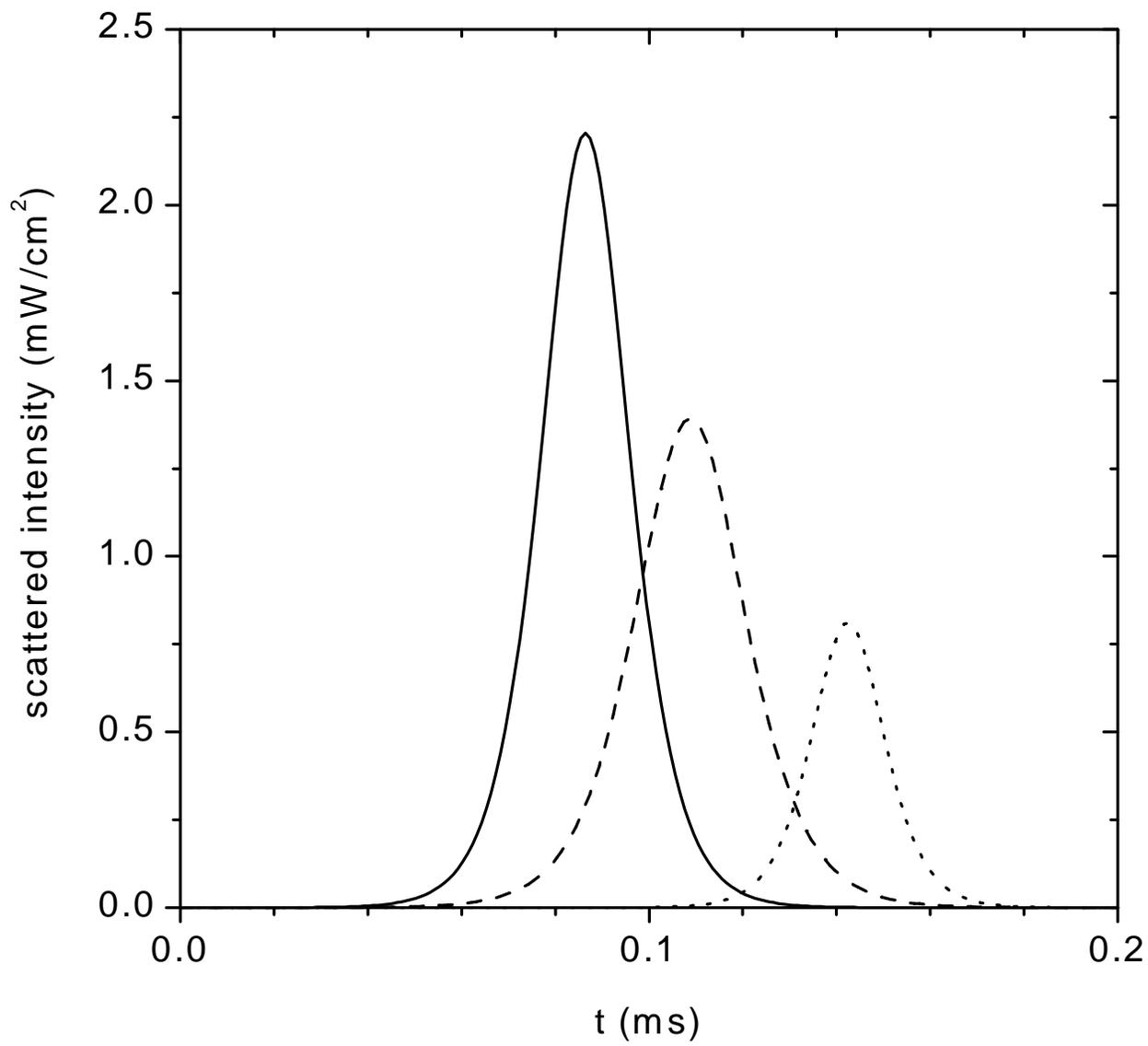
\end{document}